\documentclass[superscriptaddress,aps,pra,twocolumn,
preprintnumbers,amsmath,amssymb,floatfix,10pt]{revtex4-1}

\usepackage{graphicx}
\usepackage{dcolumn}
\usepackage{bm}
\usepackage{color}

\def\e{\epsilon}

\def\e{\mathcal{E}}

\def\ba{\begin{eqnarray}}
\def\ea{\end{eqnarray}}
\def\beq{\begin{equation}}
\def\eeq{\end{equation}}
\usepackage{subfiles}

\usepackage{hyperref}

\begin{document}

\title{Dissipative Many-body Quantum Optics in Rydberg Media}

\author{Alexey V. Gorshkov}
\affiliation{Institute for Quantum Information \& Matter, California Institute of Technology, Pasadena, CA 91125, USA}
\author{Rejish Nath}
\affiliation{Institute for Quantum Optics and Quantum Information of the Austrian Academy of Sciences, A-6020 Innsbruck, Austria}
\author{Thomas Pohl}
\affiliation{Max Planck Institute for the Physics of Complex Systems, 01187 Dresden, Germany}

\date{\today}

\begin{abstract}

We develop a theoretical framework for the dissipative propagation of quantized light in interacting optical media under  conditions of electromagnetically induced transparency (EIT). The theory allows us to determine the peculiar spatiotemporal structure of the output of two complementary Rydberg-EIT-based light-processing modules: the recently demonstrated single-photon filter and the recently proposed single-photon subtractor, which, respectively, let through and absorb a single photon. In addition to being crucial for applications of these and other optical quantum devices, the theory opens the door to the study of exotic dissipative many-body dynamics of strongly interacting photons in nonlinear nonlocal media.

\end{abstract}



\maketitle

Dissipation is commonly seen as a source of errors in quantum information and of undesired decoherence in strongly correlated many-body systems. However, recent work has shown that the very same mechanism can instead be turned into a powerful tool for quantum computing \cite{verstraete09}, self-correcting quantum memories \cite{pastawski11}, continuous quantum communication and repeaters \cite{vollbrecht11,krauter11}, manybody entangled state generation \cite{kraus08,diehl08,verstraete09,diehl10,weimer10,barreiro11,torre12}, and dissipation-driven phase transitions \cite{diehl10d}.
A particular example, realized in recent experiments \cite{dudin12,peyronel12,maxwell12}, is the propagation of quantized light fields in Rydberg media \cite{saffman10,comparat10,pohl11,low12} under the conditions of electromagnetically induced transparency (EIT) \cite{fleischhauer05}.
While Rydberg states provide strong long-range atom-atom interactions,  EIT provides strong atom-light interactions with controlled dissipation. The resulting combination gives rise to strong and often 
dissipative photon-photon interactions \cite{pritchard10,gorshkov11,sevincli11,petrosyan11}, which can be used to generate a variety of non-classical states of light \cite{lukin01b,nielsen10,honer11,olmos10,pohl10,saffman02,pedersen09,pritchard12,guerlin10,dudin12,peyronel12,dudin12b,dudin12c,maxwell12,stanojevic12,bariani12,schauss12,bariani12b} and to  implement photon-photon and atom-photon quantum gates \cite{shahmoon11,gorshkov11,petrosyan11,parigi12}. 
First wavefunction-based descriptions of two-photon propagation 
in Rydbeg EIT media have revealed the emergence of correlated two-photon losses that could enable the deterministic generation of single photons  \cite{gorshkov11,peyronel12}. Yet, the fate of the remaining photon as well as the underlying dissipative many-body dynamics have remained unclear 
despite their essential role in the performance of future Rydberg-EIT-based nonlinear optical quantum devices.

In this Letter, we address these outstanding questions and develop 
a 
theory for the dissipative many-body dynamics of quantized light fields in a strongly interacting medium. In contrast to earlier studies \cite{gorshkov11, peyronel12}, our theory provides information about the many-body density matrix of the light field, i.e.~it faithfully describes the process of populating the $n$-photon states from the $(n+1)$-photon manifold as a photon scatters.
In addition to opening the door to the study of photonic dissipative many-body physics, the theory allows one to compute the complex spatiotemporal structure of the generated non-classical light fields, whose understanding is 
 crucial for applications. 
As two important examples that illustrate this point and evince the power of our method, we consider the recently demonstrated single-photon filter \cite{gorshkov11, peyronel12} and the recently proposed single-photon subtractor \cite{honer11}. In the limit of strong interactions, our approach yields exact solutions to the dissipative many-body dynamics, provides an intuitive picture of the underlying physics, and highlights the importance of boundary effects, i.e.~of the entrance dynamics of the incoming photons.
These effects may be crucial for photon storage in a nonlinear quantum memory \cite{dudin12}, while the developed theoretical framework 
should enable the understanding of recent experiments \cite{peyronel12,maxwell12} beyond the limit of extremely weak input.

\begin{figure}[t!]
\begin{center}
\includegraphics[width = 0.99 \columnwidth]{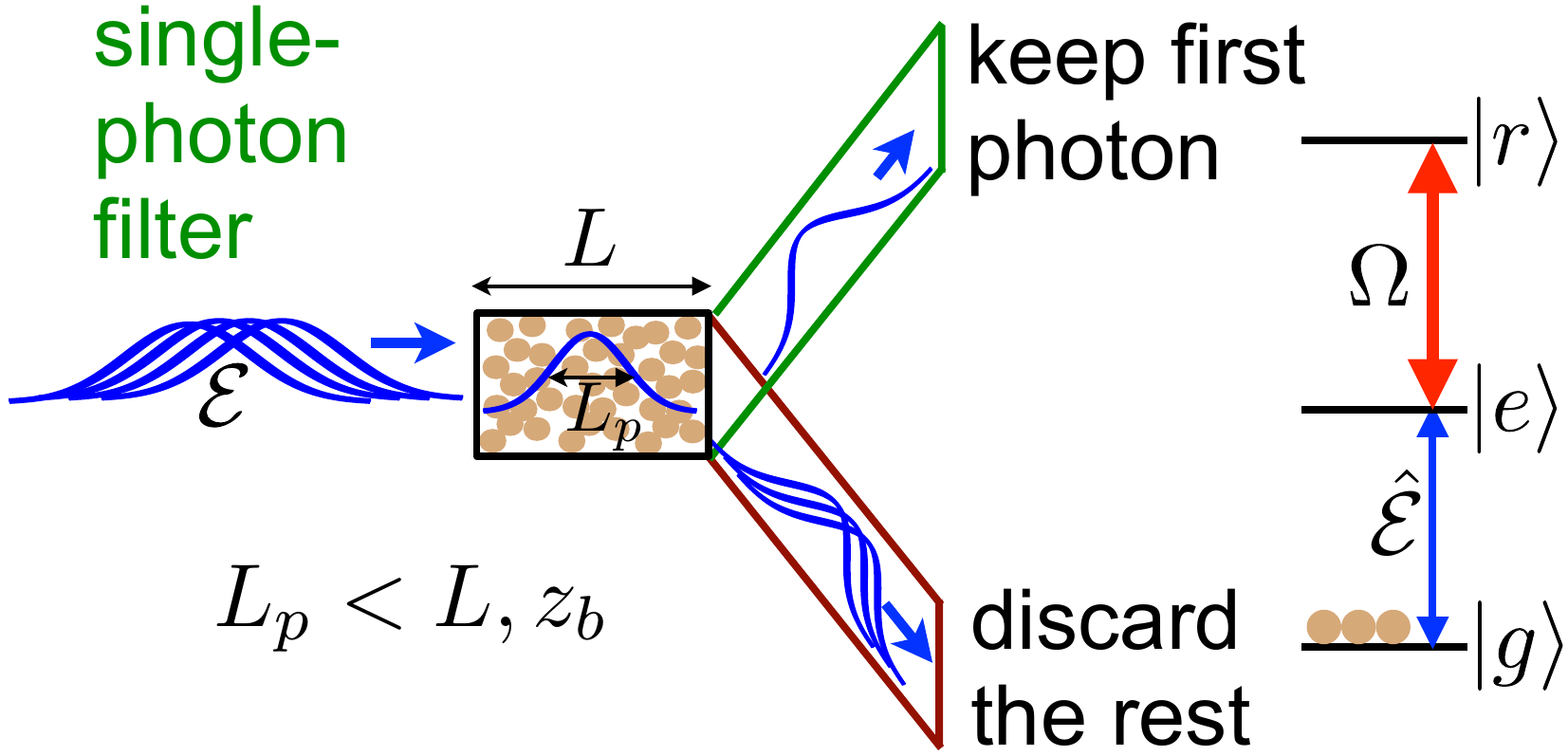}
\caption{Single-photon filter. A classical field with Rabi frequency $\Omega$ resonantly coupling the excited state $|e\rangle$ to the Rydberg state $|r\rangle$ controls the propagation of the quantum field $\hat \e$. The EIT-compressed pulse length $L_p$ is assumed to be smaller than the length $L$ of the medium and the blockade radius $z_b$.
\label{fig:scheme}}
\end{center}
\end{figure}

The basic physics can be illustrated by considering the example of a single-photon filter shown in Fig.\ \ref{fig:scheme}. In the absence of interactions, the probe field $\hat{\mathcal{E}}$ couples with an effective spin wave of Rydberg atoms $|r\rangle$ to form a slow-light polariton \cite{fleischhauer05}. 
Whenever two polaritons are within the so-called blockade radius $z_b$ of each other, 
the strong interactions between $|r\rangle$ atoms 
destroy  EIT and lead to strong dissipation  \cite{gorshkov11}.
If the EIT-compressed pulse length $L_p$ is smaller than the length $L$ of the medium and the blockade radius $z_b$, the first photon propagates without losses under EIT conditions but causes scattering of all subsequent photons 
\cite{gorshkov11}. The density matrix of the first photon is obtained by tracing over all the subsequent photons.  Since the first photon must already be inside the medium to cause scattering, the timing of the scattering events carries information about the first photon. Therefore, the transmitted single photon is impure. In the following, we develop a master-equation-type framework that allows one to determine the state of the output photons in this and other Rydberg-EIT problems, including those that do not satisfy the condition $L_p < L, z_b$.

\textit{Setup.}---Let $\hat{\e}^\dagger(z)$, $\hat{P}^\dagger(z)$, and $\hat{S}^\dagger(z)$ be the slowly-varying operators for the creation of a photon, an excitation in state $|e\rangle$, and a Rydberg excitation $|r\rangle$, respectively, at position $z$. The operators satisfy the same-time commutation relations $[\hat{\e}(z),\hat{\e}^\dagger(z')] = [\hat{P}(z),\hat{P}^\dagger(z')]= [\hat{S}(z),\hat{S}^\dagger(z')] = \delta(z-z')$. The Heisenberg equations of motion inside the medium $z \in [0,L]$  are  \cite{fleischhauer02,gorshkov07c,gorshkov11,peyronel12},
\begin{eqnarray}
\!\!\!\!\!\!\partial_t  \hat \e(z,t) &=& - \partial_z \hat \e(z,t) + i g \hat P(z,t), \label{eq:e}\\
\!\!\!\!\!\!\partial_t \hat P(z,t) &=& - \hat P(z,t) + i g \hat \e(z,t) + i \Omega \hat S(z,t), \label{eq:p}\\
\!\!\!\!\!\!\partial_t \hat S(z,t) &=& i \Omega \hat P(z,t) - i  \!\!  \int \!\! d z' V(z\!-\!z') \hat S^\dagger(z') \hat S(z') \hat S(z). \label{eq:s}
\end{eqnarray}
Here $g$ is the collective atom-photon coupling constant, $V(z) = C_6/z^6$ is the Rydberg-Rydberg interaction, time and frequencies were rescaled by $\gamma$ (the halfwidth of the $|g\rangle$-$|e\rangle$ transition), while $z$ was rescaled by $c/\gamma$. In these units, the blockade radius is given by $z_b=(C_6/\Omega^2)^{1/6}$ \cite{gorshkov11} and $c=1$.
Outside the medium ($z \notin [0,L]$), $\hat S(z,t)$ and $\hat P(z,t)$ are not defined and $(\partial_t + \partial_z) \hat \e(z,t) = 0$. 

\textit{Incoming $N$-photon Fock state.}---For simplicity, we assume that the incoming pulse is confined to a single -- for simplicity, real -- spatiotemporal mode $h(t)$ satisfying $\int d t h^2(t) = 1$. Then an incoming $N$-photon Fock state -- before entering the medium -- can be written as
\ba
|\psi(t)\rangle = \frac{1}{\sqrt{N!}} \left[ \int_{-\infty}^\infty d x h(t-x) \hat \e^\dagger(x)\right]^N |0\rangle, \label{eq:nphotoninput}
\ea
while its full density matrix at all times has the form 
\ba
\rho(t) = \sum_{n=0}^{N} \rho_n(t),  \label{eq:rhoN}
\ea
where $\rho_n$ denotes the $n$-excitation part. Since the environment possesses the information about the number of scattered photons, there are no correlations between the $N+1$ terms in Eq.\ (\ref{eq:rhoN}). The Heisenberg equations of motion (\ref{eq:e}-\ref{eq:s}) yield  master-equation-type evolution equations for the matrix elements of 
$\rho_n$. If all but at most one photon are scattered, the only nonvacuum matrix element that survives in the output field is $ee(x,y,t)  = \textrm{tr}[\rho_1(t) \hat  \e^\dagger(x) \hat \e(y)]$. As shown in the supplementary material \cite{supp}, the dissipative propagation can be solved analytically for arbitrary photon number $N$ under conditions of perfect EIT with $L_p < L, z_b$ and numerically for $N = 2$ without any restriction on the experimental parameters. In the former case, the resulting dynamics can be derived within a more general and simpler framework outlined below.

Perfect EIT with $L_p < L$ requires a large optical depth of the medium \cite{fleischhauer05}, implying that the absorption length is much smaller than the blockade radius $z_b$ and the compressed pulse length $L_p$. Since $L_p < z_b$, at most one photon can propagate through the medium without losses. Then a fundamental question directly relevant to the experiments in Refs.\ \cite{peyronel12,maxwell12} is whether all $N$ incoming photons are lost as they blockade each other's propagation or whether one photon indeed survives. 

To answer this question, we work in the Schr\"odinger picture \cite{hafezi12} and rewrite the input pulse [Eq.\ (\ref{eq:nphotoninput})] outside the medium as
\ba
|\psi(t)\rangle = \sqrt{N!} \int_{t_N >  \dots > t_1}  \left[\prod_{i=1}^N d t_i h(t_i) \e^\dagger(t-t_i)\right] |0\rangle, \label{eq:order} 
\ea
where the photons are now time-ordered. While the $N$ incoming photons are in the same spatial mode and hence indistinguishable, the possibility of time ordering is the crucial conceptual step in the derivation. As the first photon ($i = 1$) enters the medium, it turns into a Rydberg spin-wave excitation $\hat{S}$ moving at the EIT group velocity $v_g = (\Omega/g)^2 c \ll c$. 
Since $L_p < z_b$, this single Rydberg excitation turns the entire medium seen by the remaining $N-1$ photons into a resonant two-level medium. As the absorption length is much smaller than $L_p$,  all the remaining photons get scattered into some other mode $\hat{Q}$ as soon as they enter the medium. We will later trace over those loss channels, so we can assume without loss of generality that $\hat Q$ is also a one-dimensional mode with commutation relation $[\hat Q(z), \hat Q^\dagger(z')] = \delta(z-z')$ 
\cite{Note1}.
Once the entire pulse is inside the medium, we, therefore, have
\ba
\!\!\!\!\!\!|\psi(t)\rangle \!=\!\sqrt{N!} \int_{t_{N} >  \dots > t_2}  \!\! \left[\prod_{i=2}^N d t_i h(t_i) \hat Q^\dagger(t-t_i)\right] \! \! |\psi_{t_2}(t)\rangle,
\ea
where
\ba
|\psi_{t_2}(t)\rangle = - \sqrt{v_g} \int_{-\infty}^{t_2} d t_1  h(t_1) \hat S^\dagger (v_g(t-t_1)) |0\rangle.
\ea
Tracing over $\hat Q$, we obtain
\ba
&&\!\!\!\!\!\!\!\! \rho(t) =  \int d x d y ss(x,y,t) S^\dagger(y) |0\rangle \langle 0| S(x) = \label{eq:rhoNout} \\
&&\!\!\!\!\!\!\!\! \int \! d t_2 N (N-1) h^2(t_2) \! \left[\int_{t_2}^\infty  \!\!\!\! h^2(\tau) d \tau \right]^{N-2}\!\!\!\!\!\!\!\!  |\psi_{t_2}(t)\rangle \langle \psi_{t_2}(t)|,  \label{rhophysN}
\ea
where $ss(x,y,t) =\phi(x/v_g - t,y/v_g - t)/v_g$ is the density matrix of the remaining spin wave with 
\ba
\!\!\!\!\!\!\!\! \phi(x,y) = N h(-x) h(-y) \left[\int_{-\infty}^{\textrm{min}(x,y)} d z h^2(-z)\right]^{N-1}, \label{eefinalN}
\ea
which together with Eq.~(\ref{eq:rhoNout}) yields the dynamics inside the medium. 
For the purposes of this derivation, we have ignored the small photonic component $ee = v_g ss$. 
This solution has a simple physical interpretation: The trace of the integrand in Eq.~(\ref{rhophysN}) is the probability that the second photon enters the medium (and immediately scatters) in the time interval $[t_2,t_2+dt_2]$, 
while $|\psi_{t_2}(t)\rangle$ is the unnormalized spin wave that would be propagating in the medium had we detected that scattering event.

Transforming to a moving frame of reference, 
the density matrix of the output photon becomes $ee(x,y) = \phi(x,y)$ [see Eq.~(\ref{eefinalN})]. 
This result shows that exactly one photon indeed survives the dissipative entrance dynamics: $\textrm{tr}[\rho] =\int dx\phi(x,x)= 1$.  It also yields a remarkably simple result for the purity of the created photon: 
\ba
\textrm{tr}[\rho^2] =  \frac{N}{2 N - 1}.
\ea 
As expected, the purity is smaller than unity because the timing of the scattering event carries some information about the remaining spin wave $|\psi_{t_2}(t)\rangle$. Crucially for applications, the purity does not vanish but approaches $1/2$ as $N \rightarrow \infty$. Surprisingly, it is independent of the mode shape $h(t)$. Furthermore, the eigenvalues $p_i$ and eigenvectors $\phi_i(x)$ of $\phi(x,y)$ can be easily found \cite{supp} by using the change of variables $x \rightarrow \int_{-\infty}^x d z h^2(-z)$, which makes the density matrix and hence $p_i$ independent of $h(t)$.  Physically, this surprising behavior emphasizes the fact that the key role is played simply by the arrival \textit{order} of $N$ identical photons and not by the shape of the mode.   

\begin{figure}[t]
\begin{center}
\includegraphics[width = 0.99 \columnwidth]{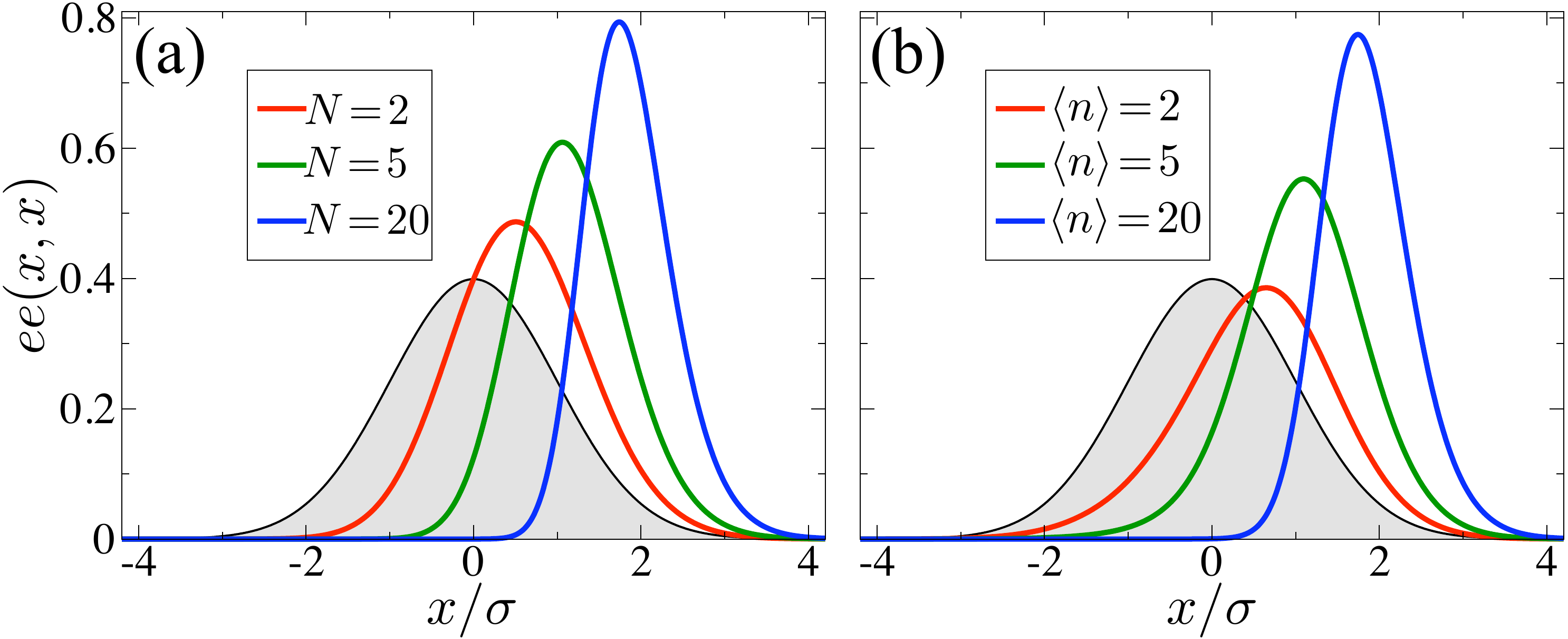}
\caption{The narrowing and advancing of the produced single-photon intensity profile $ee(x,x)$ with increasing input intensity. The input mode $h^2(x) \propto \exp[-x^2/(2 \sigma^2)]$ is shaded. (a) Input is a Fock state with the indicated value of $N$. 
(b) Input is a coherent  state with the indicated value of $\langle n \rangle$. \label{fig:out}}
\end{center}
\end{figure}

This dynamics at the medium boundary leads to a slight narrowing and advancing of the single-photon pulse $\phi(x,x)$, as shown in Fig.\ \ref{fig:out}(a) for a typical Gaussian input mode. 
This behavior can be traced back to the first scattering event, which projects the leading photon into the medium. This effect becomes more pronounced with increasing $N$, since the larger $N$ is the sooner the first scattering event takes place. 
More succinctly, the probability distribution of the \textit{first photon} is obviously advanced and  narrower relative to the normalized probability distribution $h^2(t)$ of  the \textit{entire incident pulse}. 
Fortunately, $\phi(x,x)$ and $\phi_i(x)$ shorten extremely slowly with $N$ as $\sim 1/\sqrt{\log N}$, keeping the associated  losses at a minimum.  

 \begin{figure}[t]
\begin{center}
\includegraphics[width = 0.99 \columnwidth]{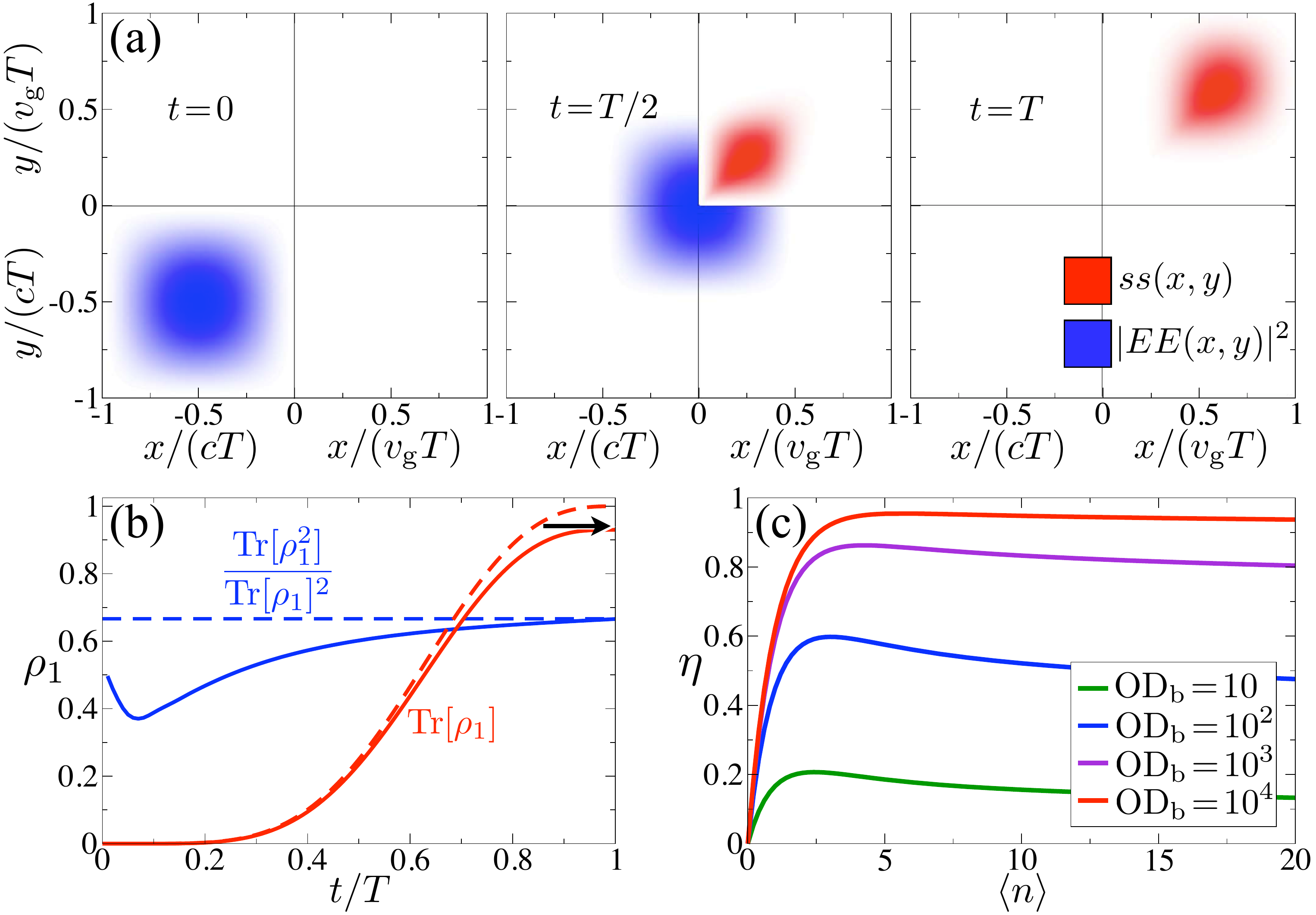}
\caption{ (a) For an incoming $N = 2$ Fock state, the two-photon amplitude and the single-spin-wave density matrix at different times. 
(b) Approximate analytical (dashed) and exact numerical (solid) results for the efficiency $\textrm{tr}[\rho_1]$ (red) and purity $\textrm{tr}[\rho_1^2]/\textrm{tr}[\rho_1]^2$ (blue) of the single excitation. The purple arrow indicates the efficiency of retrieving Eq.\ (\ref{eq:rhoNout}) with $N = 2$.  (c) As a function of $\langle n \rangle$ of the incoming coherent pulse, for the indicated values of the blockaded optical depth $\textrm{OD}_\textrm{b}$, the estimated efficiency $\eta$ of the single-photon source. 
\label{fig:numerics}}
\end{center}
\end{figure}

To verify this intuitive picture, we carried out numerical simulations for $N = 2$ incoming photons and $h(t) \propto 1- 4 (t/T - 0.5)^2$  ($t \in [0,T]$) using the full propagation equations derived from Eqs.\ (\ref{eq:e}-\ref{eq:s}) \cite{supp}. This form of $h(t)$ is motivated by optimal photon storage \cite{gorshkov07,gorshkov07c}. The results are shown in Fig.\ \ref{fig:numerics}(a,b). In Fig.\ \ref{fig:numerics}(a), the left-bottom quadrant corresponds to both photons  being still outside the medium, so $t = 0$ is described by Eqs.\ (\ref{eq:nphotoninput},\ref{eq:order}). The top-right quadrant corresponds to both photons being inside the medium, so $t = T$ is described by Eq.\ (\ref{eq:rhoNout}). Finally the remaining two quadrants correspond  to the first photon being already inside the medium while the second photon is still outside. Fig.\ \ref{fig:numerics}(b) shows a comparison to the analytical prediction from Eqs.~(\ref{eq:rhoNout},\ref{eefinalN}) \cite{Note2}.
While imperfections keep the single-photon conversion efficiency slightly away from unity, the overall physical picture is very well confirmed by our numerical simulations. To verify that losses induced by the finite width of the EIT transparency window -- and not the correlated photon dynamics upon pulse entrance -- constitute the dominant imperfection, the purple arrow in Fig.\ \ref{fig:numerics}(b) indicates the efficiency of retrieving Eq.\ (\ref{eq:rhoNout}) with $N=2$ back out of the medium. 

\textit{Arbitrary incoming state.}---
Since any mixed state can be represented as a classical mixture of pure states, it is sufficient to consider an arbitrary pure input state
\ba
|\psi\rangle = \sum_m c_m |m\rangle \label{eq:genpur}.
\ea
Following the above treatment, we trace over all photons except for the first one to obtain 
\ba
\rho = (c_0 |0\rangle + c_1 |1\rangle) (c_0^* \langle 0| + c_1^* \langle 1|) + \sum_{m \geq 2} |c_m|^2 \rho_m, \label{eq:rhofull}
\ea
where $\rho_m$ is the single photon obtained from $|m\rangle$. The single-photon conversion efficiency is $1- |c_0|^2$, i.e. limited by the vacuum component of the input state. The corresponding purity is $\textrm{tr}[\rho_1^2]/(1- |c_0|^2)^2$, where
\ba
\textrm{tr}[\rho_1^2] =  \sum_{m,n \geq 1}  |c_n|^2 |c_m|^2 \frac{2 m n}{(n+m-1)(n+m)}. \label{rhosqfull}
\ea
For a coherent input, $|c_i|^2 = e^{-\langle n\rangle} \langle n \rangle^i/i !$, the efficiency is thus $1-e^{-\langle n\rangle}$, while the single-photon purity is $(1- e^{-\langle n\rangle})^{-2} (1 - e^{-2 \langle n \rangle} (1 + 2 \langle n \rangle)) /2$, which falls off monotonically from $1$ to $1/2$ with increasing $\langle n \rangle$. Since $|c_0|^2$ drops exponentially with $\langle n \rangle$, a small average number of incoming photons $\langle n \rangle \sim 10$ is sufficient to make the single-photon source deterministic. Repeating the above derivations, one obtains for the output density matrix
\ba
\!\!\!\!\!\phi(x,y) = \langle n \rangle h(-x) h(-y) \exp\left[\!- \langle n \rangle \!\! \int^{\infty}_{\textrm{min}(x, y)} \!\!\!\!\!\! h^2(-z) d z \right], \label{eefinaln}
\ea
which can easily be diagonalized \cite{supp}. As for a Fock-state input with Gaussian $h(x)$,  
the output pulse shortens extremely slowly with increasing $\langle n \rangle$ ($ \sim 1/\sqrt{\log \langle n \rangle}$), as shown in Fig.\ \ref{fig:out}(b). 

The efficiency of  this single-photon source -- imperfect due to the finite width of the EIT window -- can be estimated from the analytical form of the density matrix without involving interactions.  
We assume that the incoming pulse is stored without interactions into the spin-wave $s(z) \propto1-4(z/L-0.5)^2$ 
and that the single-spin-wave density matrix [Eq.~(\ref{eefinalN}) or Eq.~(\ref{eefinaln})] is  retrieved forward. 
The efficiency $\eta$ of the single-photon source can then be estimated as the product of these two -- storage and retrieval --  efficiencies. 
The $\langle n\rangle$-dependence of $\eta$ for a coherent input is shown in Fig.\ \ref{fig:numerics}(c)  for 
 different blockaded optical depths OD$_b$, assuming the entire medium is blockaded. 
 The relatively poor scaling of the efficiency with OD$_b$ results from the cusp of the density matrix $\phi(x,y)$ along the diagonal ($x=y$), which carries high-frequency components. In a magneto-optical trap (density $\mathcal{N} \sim 10^{12} \textrm{ cm}^{-3}$), OD$_\textrm{b}\sim 10$ \cite{peyronel12}, and hence $\eta \approx 0.2$ can be achieved. In a BEC  \cite{viteau11b,schauss12}, $\mathcal{N} \sim 10^{14} \textrm{ cm}^{-3}$ can give OD$_\textrm{b} \sim 1000$ and $\eta \approx 0.9$. The efficiency can be further increased by using photonic waveguides \cite{christensen08,bajcsy09,vetsch10,goban12} and by further optimizing $h(t)$ and retrieving backwards \cite{gorshkov07,gorshkov07c}. 

Despite their impurity, the single photons produced with this method are a valuable resource. In particular, the impurity would not interfere with applications that do not rely on Hong-Ou-Mandel interference \cite{hong87}, such as optical quantum computing with impurity-insensitive two-qubit gates (e.g.~\cite{gorshkov10b}) or the BB84 quantum key distribution protocol \cite{bennett84}. For applications that rely on Hong-Ou-Mandel interference, the photon can be purified in the following ways. First, the detection of the first scattered photon would yield a pure photon. Second, the impure single photon can be purified to the dominant eigenvector $|\phi_1\rangle$ with probability $p_1$ ($p_1 = 0.69$ for $\langle n \rangle \gg 1$). 
This can be accomplished, e.g., using an atomic ensemble in a cavity \cite{simon07c} to store only the mode $|\phi_1\rangle$, heralded by the absence of a click at the cavity output, followed by retrieval, which can be shaped into any desired mode \cite{gorshkov07b}. 

\textit{Photon subtraction.}---To demonstrate the versatility of the developed theory, we now apply it to the Rydberg-based single-photon subtractor proposed in Ref.~\cite{honer11}, showing that this scheme also yields impure output states. The detailed physics of such a setting \cite{supp} is complementary to the single-photon source in so far as the density matrix of the remaining photons is obtained by tracing out the first one. Since the timing of the absorption carries information about the remaining photons, the density matrix of the latter is impure as well. In fact, the single-photon subtractor and the single-photon filter complement each other to make the original pure state. Hence, the impurity and the entire eigenspectrum of the reduced density matrix are identical in the two cases.  

This can be shown by tracing over the first photon in Eq.\ (\ref{eq:genpur}) to obtain
\ba
\rho = |c_0|^2 |0\rangle \langle 0| + \int_{-\infty}^\infty d t_1 h^2(t_1) |\psi_{t_1}(t)\rangle \langle \psi_{t_1}(t)|, \label{eq:subtract}
\ea
where
\ba
|\psi_{t_1}(t)\rangle = \sum_{m \geq 1} c_m \frac{\sqrt{m!}}{(m-1)!} \left[\int_{t_1}^\infty d t' h(t') \e^\dagger(t-t')\right]^{m-1} |0\rangle,\nonumber
\ea
which has the same eigenspectrum as Eqs.\  (\ref{eq:rhofull}). 

\textit{Outlook.}---In conclusion, we extended the dynamics of open quantum systems of Rydberg atoms \cite{lee12,lee12c,ates11,olmos11,hoening12,glaetzle12,gunter12,robert-de-saint-vincent12,nipper12,pupillo10} to include the dissipative quantum dynamics of the propagating light field, which is crucial for the understanding of recent experiments \cite{peyronel12,maxwell12}. While we focused on the case of blockade radius that is larger than the size of the EIT-compressed pulse, the developed framework also applies to finite blockade radii \cite{peyronel12,maxwell12} and can be easily extended to a time-dependent blockade radius, as relevant for photon storage via time-dependent control fields.
Extensions to non-dissipative unitary evolution \cite{parigi12}, media with longitudinal density variations, incomplete transverse blockade, as well as finite Rydberg-state lifetime are straightforward \cite{peyronel12}. Finally, we expect our calculations to be extendable to other light-processing modules, 
such as the quantum filter of Refs.\ \cite{taichenachev05b,taichenachev08b}. Most importantly, our approach may lead to a simplified effective theory for the many-body dissipative dynamics of correlated photons in strongly interacting media.

We thank M.\ Lukin, V.\ Vuleti\'{c}, O.\ Firstenberg, T.\ Peyronel, Q.-Y.\ Liang, J.\ Preskill, M.\ Fleischhauer, J.\ Otterbach, D.\ Petrosyan, J.\ Honer, T.\ Lee, S.\ Hofferberth, T.\ Pfau, P.\ Grangier, and C.\ Adams    
for discussions. This work was supported by NSF, IQIM, 
 the Lee A. DuBridge and Gordon and Betty Moore foundations, CUA and the EU through the Marie Curie ITN COHERENCE. 


%

\clearpage

\title{Supplementary online material to the manuscript:\\``Dissipative Many-body Quantum Optics in Rydberg Media"}

\author{Alexey V. Gorshkov}
\affiliation{Institute for Quantum Information \& Matter, California Institute of Technology, Pasadena, CA 91125, USA}
\author{Rejish Nath}
\affiliation{Institute for Quantum Optics and Quantum Information of the Austrian Academy of Sciences, A-6020 Innsbruck, Austria}
\author{Thomas Pohl}
\affiliation{Max Planck Institute for the Physics of Complex Systems, 01187 Dresden, Germany}

\maketitle


\section{Equations of motion for two incoming photons} 

In this Section, we give the equations of motion that were used to 
do the numerics for two incoming photons and to obtain Fig.~3(a,b) in the main text.

In the case of two incoming photons, the full 
density matrix 
\ba
\rho(t) &=& \epsilon(t) |0\rangle \langle 0| + \rho_1(t) + |\psi_2(t)\rangle \langle \psi_2(t)| \label{XXeq:rhoP}
\ea
consists of the unnormalized two-excitation wavefunction
\ba
\!\!\!\! |\psi_2(t)\rangle & = &  \frac{1}{2} \int d x \int d y EE(x,y,t) \hat \e^\dagger(x) \hat \e^\dagger(y) |0\rangle \nonumber \\
&&+ \int d x \int' d y  EP(x,y,t) \hat \e^\dagger(x) \hat P^\dagger(y) |0\rangle \nonumber \\
&&+ \int d x \int' d y  ES(x,y,t) \hat \e^\dagger(x) \hat S^\dagger(y) |0\rangle \nonumber \\
&& + \frac{1}{2} \int' d x \int' d y PP(x,y,t) \hat P^\dagger(x) \hat P^\dagger(y) |0\rangle \nonumber \\
&&+ \int' d x \int' d y  PS(x,y,t) \hat P^\dagger(x) \hat S^\dagger(y) |0\rangle \nonumber \\
&& + \frac{1}{2} \int' d x \int' d y SS(x,y,t) \hat S^\dagger(x) \hat S^\dagger(y) |0\rangle,
\ea
the unnormalized single-excitation density matrix 
\ba
\rho_1(t) &=&  \int d x \int d y \, ee(x,y,t) \hat \e^\dagger(y) |0\rangle \langle 0| \hat \e(x) \nonumber \\
&& + \int d x \int' d y  \,  ep(x,y,t) \hat P^\dagger(y) |0\rangle \langle 0| \hat \e(x) \nonumber \\
&& +  \int' d x \int d y  \, pe(x,y,t) \hat \e^\dagger(y) |0\rangle \langle 0| \hat P(x) \nonumber \\ 
&& + \int d x \int' d y  \, es(x,y,t) \hat S^\dagger(y) |0\rangle \langle 0| \hat \e(x) \nonumber \\
&& +  \int' d x \int d y  \, se(x,y,t) \hat \e^\dagger(y) |0\rangle \langle 0| \hat S(x) \nonumber \\
&&+  \int' d x \int' d y  \, pp(x,y,t) \hat P^\dagger(y) |0\rangle \langle 0| \hat P(x) \nonumber \\
&& + \int' d x \int' d y  \, ps(x,y,t) \hat S^\dagger(y) |0\rangle \langle 0| \hat P(x) \nonumber \\
&& +  \int' d x \int' d y  \, sp(x,y,t) \hat P^\dagger(y) |0\rangle \langle 0| \hat S(x) \nonumber \\
&&+  \int' d x \int' d y  \, ss(x,y,t) \hat S^\dagger(y) |0\rangle \langle 0| \hat S(x),  \label{Xrho1}
\ea
and the vacuum component $\epsilon(t) |0\rangle \langle 0|$. Here $\int$ integrates over $(-\infty,\infty)$, while $\int'$ integrates over $[0,L]$. Without loss of generality, we take $EE$, $PP$, and $SS$ to be symmetric [e.g.~$EE(x,y) = EE(y,x)$]. 
If the input state had correlations between  different Fock states, one would need to include coherences between manifolds of different photon number; the method we discuss  can be naturally generalized to these situations.

All terms in $\rho(t=0)$ vanish except for $EE(x,y,0) = \sqrt{2} h(-x) h(-y)$, where we assume 
$h(t < 0) = 0$.
The equations of motion for $EE$, $EP$, $ES$, $PP$, $PS$, and $SS$ can be obtained by expressing them in terms of $|\Psi_2\rangle$ [e.g.\ $ES(x,y) = \langle 0|\hat \e(x) \hat S(y)|\psi_2(t)\rangle$] and using Eqs.\ (1-3) in the main text.
For $x \notin [0,L]$, $y \in [0,L]$, they are
\ba
(\partial_t + \partial_x + \partial_y) EE &=& i g EP, \label{Xeq:EE1}\\
(\partial_t + \partial_x + 1) EP &=& i g EE + i \Omega ES,\\
(\partial_t + \partial_x) ES &=& i \Omega EP,\label{Xeq:ES1}
\ea
and describe 
the EIT propagation of photon $y$, while photon $x$ propagates outside the medium with the speed of light ($c=1$ in our units). Using $EE(x,y,t) =  \sqrt{2} h(t-x) h(t-y)$ to set the boundary conditions at $y = 0$, these equations are solved for $x \leq 0$, $y \in [0,L]$ to give the boundary conditions for the equations in the region $x,y \in [0,L]$:
 \ba
 (\partial_t+  \partial_x +  \partial_y) EE &=&  i g EP_+,\\
 (\partial_t + \partial_x + 1) EP &=& i g (EE + PP) +  i \Omega ES,\\
  (\partial_t + \partial_x) ES &=& i g PS + i \Omega EP, \label{Xeq:ES2}\\
 (\partial_t + 2) PP &=& i g EP_+ + i \Omega ES_+,\\
 (\partial_t + 1) PS &=& i g ES + i \Omega (PP + SS), \label{Xeq:PS2}\\
 (\partial_t + i V(r)) SS &=& i \Omega PS_+,
\ea
where $EP_\pm(x,y) = EP(x,y) \pm EP(y,x)$,  $ES_\pm(x,y) = ES(x,y) \pm ES(y,x)$, $PS_\pm(x,y) = PS(x,y) \pm PS(y,x)$, and $r = x-y$. The solution to these equations can then be used to set the boundary conditions at $x = L$ for Eqs.\ (\ref{Xeq:EE1}-\ref{Xeq:ES1}) in the region $x \geq L$, $y \in [0,L]$, which can in turn be used  calculate the outgoing two-photon pulse.

Now we turn to the evolution equations for 
the single-excitation density matrix $\rho_1$. We first note that
\ba
&&es(x,y) = \langle \hat \e^\dagger(x) \hat S(y)\rangle - \int d z EE^*(x,z) ES(z,y) \label{Xeq:esbar} \\
&& - \int' d z EP^*(x,z) PS(z,y) - \int' d z ES^*(x,z) SS(z,y). \nonumber
\ea
The equation of motion for $\langle \hat \e^\dagger(x) \hat S(y)\rangle$ follows from Eqs.\ (1-3) in the main text. Together with the equations of motion for the two-photon amplitudes, this yields the equation of motion for $es(x,y)$, and, similarly, for all matrix elements of $\rho_1$. The following source terms will describe the transfer of population from $|\psi_2\rangle$ to $\rho_1$:
\ba
f_{ee}(x,y) &=& 2 \int' dz EP^*(x,z) EP(y,z),\\
f_{ep}(x,y) &=& 2 \int' dz EP^*(x,z) PP(y,z),\\
f_{es}(x,y) &=& 2 \int' dz EP^*(x,z) PS(z,y),\\
f_{pp}(x,y) &=& 2 \int' dz PP^*(x,z) PP(y,z),\\
f_{ps}(x,y) &=& 2 \int' dz PP^*(x,z) PS(z,y),\\
f_{ss}(x,y) &=& 2 \int' dz PS^*(z,x) PS(z,y).\label{Xeq:fss}
\ea
As expected, in the interaction-free case ($V = 0$) and assuming perfect EIT, the source terms vanish because $|e\rangle$ is never populated, so all components of $|\psi_2\rangle$ involving $P$ vanish. With these definitions, 
for $x,y \in [0,L]$,
\ba
\!\!\!\!\!\!\!\!\!\!\!\!(\partial_t  + \partial_x + \partial_y) ee &=& i g (ep-pe) + f_{ee},\label{Xeq:ee3}\\
\!\!\!\!\!\!\!\!\!\!\!\!(\partial_t +  \partial_x + 1) ep &=&  i g  (ee - pp) +i \Omega es +  f_{ep}, \\
\!\!\!\!\!\!\!\!\!\!\!\!(\partial_t +  \partial_x)  es &=&  i \Omega ep - i g \, ps + f_{es},\label{Xeq:es3}\\
\!\!\!\!\!\!\!\!\!\!\!\!(\partial_t +  2) pp &=&  i g (pe-ep) + i \Omega (ps-sp) + f_{pp},\\
\!\!\!\!\!\!\!\!\!\!\!\!(\partial_t +  1) ps &=&  - i g \, es + i \Omega(pp-ss) + f_{ps},\label{Xeq:ps3}\\
\!\!\!\!\!\!\!\!\!\!\!\!\partial_t ss &=&  i \Omega (sp-ps)+ f_{ss},\label{Xeq:ss3}
\ea
while $pe(x,y) = ep^*(y,x)$, $se(x,y) = es^*(y,x)$, and $sp(x,y) = ps^*(y,x)$. Equations of motion outside of $x,y \in [0,L]$ can be obtained in the same way.

We do numerical calculations in the regime of good EIT ($\lesssim 1\%$ single-photon loss). Thus, to a good approximation, photon scattering occurs only when both photons are inside the medium. 
We therefore 
solve Eqs.\ (\ref{Xeq:ee3}-\ref{Xeq:ss3}) with vanishing initial and boundary conditions.

We note that the equations can easily be extended \cite{Xpeyronel12} to include longitudinally varying density, finite decoherence rate of $S$, as well as cases where the blockade radius is smaller than the transverse extent of the probe beam. 

\section{Ideal single-photon generation from 2 photons \label{Xsec:2}}

In this Section, in the case of an input Fock state with $N = 2$, we show how Eqs.\ (9,11) in the main text arise from the full equations of motion presented above.

Let's assume 
that 
EIT is perfect and that  $L_p  < L < z_b$. 
Then, for $x \leq 0$ and $y \in [0,L]$, Eqs.\ (\ref{Xeq:EE1}-\ref{Xeq:ES1}) give
\ba
ES(x,y,t) = - \sqrt{2/v_g} h(t-x) h(t- y/v_g),
\ea
where $v_g = (\Omega/g)^2$ in our units. From Eqs.\ (\ref{Xeq:ES2},\ref{Xeq:PS2}), we obtain $\partial_t ES \approx - \partial_x ES - g^2 ES$, so that, 
for $x,y \in [0,L]$,  
\ba
PS(x,y,t) & \approx & i g ES(x,y,t) \approx i g ES(x=0,y,t-x) e^{-g^2 x} \nonumber \\
&\approx & - i g \sqrt{2/v_g} h(t) h(t - y/v_g) e^{- g^2 x}, \label{XESq1}
\ea
which describes the absorption of the two-excitation amplitude over the absorption length $1/g^2 \ll L_p$. 
Inserting this expression into Eq.\ (\ref{Xeq:fss}), we obtain
\ba
f_{ss}(x,y,t) = 2 h^2(t) h(t - x/v_g) h(t- y/v_g)/v_g. \label{Xeq:fss2}
\ea
Then from Eqs.\ (\ref{Xeq:es3},\ref{Xeq:ps3},\ref{Xeq:ss3}), 
to a good approximation,
 \ba
 es &=& - (\Omega/g) ss + (\Omega/g^3) \partial_x ss,\label{Xeq:esapprox}\\
\partial_t ss &=& - 2 \Omega^2 ss - \Omega g (se + es) + f_{ss}.\label{Xeq:sstemp}
 \ea
 Inserting Eq.\ (\ref{Xeq:esapprox}) into Eq.\ (\ref{Xeq:sstemp}), we obtain
\ba
\partial_t ss = - v_g \partial_R ss + f_{ss}(x,y,t),\label{Xssvg}
\ea
which describes how the source $f_{ss}$ puts excitations into $ss$; and, as soon as the excitations are put in, they start moving at $v_g$ along $R = (x+y)/2$. This equation assumes the wavepacket's frequency components all fit inside the EIT transparency window. Solving this equation gives
\ba
\!\!\!\!\!\!\!\!\!\!\!\! ss(x,y,t) = 
 \frac{2}{v_g} h(t \!-\! \frac{x}{v_g}) h(t \!-\! \frac{y}{v_g}) \int_{t - \frac{\textrm{\scriptsize{min}}(x,y)}{v_g}}^t \!\!\!\! d t' h^2(t'), \label{Xssfinal}
\ea 
which, for $N = 2$, generalizes Eqs.\ (9,11) in the main text to cases when the pulse has only partially entered the medium. Eq.\ (\ref{Xssfinal}) yields $\textrm{Tr}[\rho_1^2]/\textrm{Tr}[\rho_1]^2 = 2/3$ for all $t$ and $\textrm{Tr}[\rho_1] = \left[\int^t_{-\infty} d \tau h^2(\tau)\right]^2$, which give the dashed lines in Fig.\ 3(b) in the main text.
To a good approximation, $\rho_1$ satisfies the dark-state-polariton condition $ee = - \sqrt{v_g} es = v_g ss$. This derivation can easily be extended to include the effect of a finite EIT transparency window width, which partially explains the slight discrepancy between analytics and numerics in Fig.\ 3(b) in the main text. 

\section{Ideal single-photon generation from $N$ photons}

In this Section, we generalize Sec.~\ref{Xsec:2} to arbitrary $N$.

Let $\mathbf{x}_m \equiv x_1, \dots, x_m$, $E_m \equiv E \dots E$ (where $E$ is repeated $m$ times to denote the $m$-photon wavefunction), and $h(t-\mathbf{x}_m) \equiv \prod_{i=1}^m h(t-x_i)$. Then, for $\mathbf{x}_N < 0$, the incoming $N$-photon state is given by 
\ba
E_N(\mathbf{x}_N) = \sqrt{N!} h(t-\mathbf{x}_N).
\ea
Once the first two photons enter the medium ($\mathbf{x}_{N-2} < 0$ and $x_{N-1},x_N > 0$), we have, in analogy with Eq.~(\ref{XESq1}),
\ba
&&E_{N-2}PS(\mathbf{x}_N)  \\
&& =  - i g \sqrt{N!/v_g} h(t-\mathbf{x}_{N-2}) h(t) h(t - x_N/v_g) e^{- g^2 x_{N-1}}. \nonumber
\ea
So, by analogy with Eqs.~(\ref{Xeq:fss},\ref{Xeq:fss2}), 
\ba
&&f_{e_{N-2}se_{N-2}s}(\mathbf{x}_{N-1},\mathbf{x}_{N-1}')  \\ 
&& = 2  \!\! \int'\!\!\!\! d z E_{N-2} PS^*(\mathbf{x}_{N-2}, z, x_{N-1}) E_{N-2} PS(\mathbf{x}'_{N-2}, z, x'_{N-1}) \nonumber  \\
&&=  \frac{N!}{v_g} h(t \!-\! \mathbf{x}_{N-2}) h(t\!-\!\mathbf{x}'_{N-2})  h^2(t)  h(t \!-\! \frac{x_{N-1}}{v_g}) h(t\! -\! \frac{x'_{N-1}}{v_g}). \nonumber
\ea
Applying group velocity propagation along $(x_{N-1}+x'_{N-1})/2$ [as in Eq.~(\ref{Xssvg})], we have [as in Eq.~(\ref{Xssfinal})]
\ba
&&e_{N-2}se_{N-2}s(\mathbf{x}_{N-1},\mathbf{x}_{N-1}') = \frac{N!}{v_g}  h(t - \mathbf{x}_{N-2}) h(t-\mathbf{x}_{N-2})   \nonumber  \\
&& \times   h(t \!-\! \frac{x_{N-1}}{v_g}) h(t \!-\! \frac{x'_{N-1}}{v_g})  \int_{t - \frac{\textrm{\scriptsize{min}}(x_{N-1},x'_{N-1})}{v_g}}^t \!\! d t' h^2(t').
\ea
Allowing now the third photon to enter the medium ($x_{N-2}, x'_{N-2} > 0$), we have
\ba
&&f_{e_{N-3}se_{N-3}s}(\mathbf{x}_{N-2},\mathbf{x}_{N-2}')   \\
&&= e_{N-2}se_{N-2}s(\mathbf{x}_{N-3},0,x_{N-2},\mathbf{x}_{N-3}',0,x_{N-2}')  \nonumber \\
&& = \frac{N!}{v_g} h(t - \mathbf{x}_{N-3}) h(t-\mathbf{x}'_{N-3})  h^2(t)  \nonumber \\
&& \times h(t - \frac{x_{N-2}}{v_g}) h(t - \frac{x'_{N-2}}{v_g})  \int_{t - \frac{\textrm{\scriptsize{min}}(x_{N-2},x'_{N-2})}{v_g}}^t d t' h^2(t'). \nonumber 
\ea
Applying group velocity propagation along $(x_{N-2} + x'_{N-2})/2$, we have
\ba
&&e_{N-3}se_{N-3}s(\mathbf{x}_{N-2},\mathbf{x}_{N-2}')  \nonumber \\ 
&& = \frac{N!}{v_g} h(t - \textbf{x}_{N-3}) h(t-\textbf{x}'_{N-3})   h(t - \frac{x_{N-2}}{v_g}) h(t - \frac{x'_{N-2}}{v_g}) \nonumber \\
&&\times \frac{1}{2} \left[\int_{t - \frac{\textrm{\scriptsize{min}}(x_{N-2},x'_{N-2})}{v_g}}^t d t' h^2(t')\right]^2.
\ea
Allowing the fourth photon to enter the medium, we have
\ba
&&f_{e_{N-4}se_{N-4}s}(\mathbf{x}_{N-3},\mathbf{x}_{N-3}') \\
&& = e_{N-3}se_{N-3}s(\mathbf{x}_{N-4},0,x_{N-3},\mathbf{x}_{N-4}',0,x_{N-3}')   \nonumber \\
&& = \frac{N!}{v_g} h(t - \mathbf{x}_{N-4})h(t-\mathbf{x}'_{N-4})  h^2(t) \nonumber \\
&& \times h(t\! -\! \frac{x_{N-3}}{v_g}) h(t \!-\! \frac{x'_{N-3}}{v_g}) \frac{1}{2}\!\! \left[\int_{t - \frac{\textrm{\scriptsize{min}}(x_{N-3},x'_{N-3})}{v_g}}^t \!\! d t' h^2(t')\right]^2. \nonumber
\ea
Applying group velocity propagation along $(x_{N-3} + x'_{N-3})/2$, we have
\ba
&&e_{N-4}se_{N-4}s(\mathbf{x}_{N-3},\mathbf{x}_{N-3}') \nonumber  \\ 
&& = \frac{N!}{v_g} h(t - x_1) \dots h(t-x'_{N-4})   h(t - \frac{x_{N-3}}{v_g}) h(t - \frac{x'_{N-3}}{v_g}) \nonumber \\
&& \times \frac{1}{3!} \left[\int_{t - \frac{\textrm{\scriptsize{min}}(x_{N-3},x'_{N-3})}{v_g}}^t d t' h^2(t')\right]^3.
\ea
We continue in this way until we reach
\ba
&&ss(x_1,x_1') =  
\nonumber \\
&&=  \frac{N}{v_g}   h(t - \frac{x_1}{v_g}) h(t - \frac{x'_1}{v_g}) \left[\int_{t - \frac{\textrm{\scriptsize{min}}(x_1,x'_1)}{v_g}}^t d t' h^2(t')\right]^{N-1},
\ea
which generalizes Eqs.\ (9,11) in the main text to cases when the pulse has only partially entered the medium. 

\section{Eigenvectors of the single-photon density matrix}

In this Section, we study the eigenvectors and eigenvalues of the single-photon density matrix, Eqs.~(11) and (16) in the main text, obtained via single-photon filtering  from Fock-state and  coherent-state inputs, respectively.



We first study the eigenvectors $\phi_i$ and eigenvalues $p_i$ of the single-photon density matrix 
$\phi(x,y)$ given in Eq.~(11) in the main text.
Defining $\tilde x = \int_{-\infty}^x d z h^2(-z)$, 
we obtain $\rho =  \int_0^1 d \tilde x d \tilde y  \tilde \phi(\tilde x, \tilde y)  \hat{\tilde \e}^\dagger(\tilde y) |0\rangle \langle0| \hat{\tilde \e}(\tilde x)$,  where $\tilde \phi(\tilde x, \tilde y) =  N \left[\textrm{min}(\tilde x, \tilde y)\right]^{N-1}$, $ \hat {\tilde \e}(\tilde x) = \hat \e(x)/h(-x)$, $[\hat{\tilde \e} (\tilde x), \hat{\tilde \e}^\dagger(\tilde y)] = \delta(\tilde x - \tilde y)$. The eigenvalues $p_i$ are then the solutions of the characteristic equation $J_{-1/N}\left[2 \sqrt{(N - 1)/(N p)}\right] = 0$. In particular, in the limit $N \rightarrow \infty$, $p_i$ are the roots of $J_0[2/\sqrt{p}] = 0$. The eigenvectors of $\tilde \phi(\tilde x, \tilde y)$ are
$\tilde \phi_i(\tilde x) \propto \tilde x^{(N-1)/2} J_{1-1/N}\left[2 \sqrt{(N - 1)/(N p_i)} \tilde x^{N/2}\right]$.
In particular, for $N = 2$, $p_i = 2 \pi^{-2} \left(n-\frac{1}{2}\right)^{-2}$, $\tilde \phi_i(\tilde x) = \sqrt{2}  \sin\left[\pi \left(n - \frac{1}{2}\right) \tilde x \right]$. While $\tilde \phi(\tilde x,\tilde x)$ and $\tilde \phi_i(\tilde x)$ shorten as $1/N$ with increasing $N$,   $\phi(x,x)$ and 
$\phi_i(x) = h(-x) \tilde \phi_i\left(\tilde x \right)$
shorten much slower as $1/\sqrt{\log N}$ for a Gaussian $h(x)$. 

We now study the eigenvectors of the single-photon density matrix $\phi(x,y)$ given in Eq.~(16) in the main text.
Following the same change of variables, we obtain $\tilde \phi(\tilde x, \tilde y) = \langle n \rangle \exp\left[- \langle n \rangle (1- \textrm{min}(\tilde x, \tilde y))\right]$, which, for $\langle n \rangle \gg 1$, agrees with the above $\phi(\tilde x, \tilde y)$ provided one identifies $N$ with $\langle n\rangle$. For general $\langle n \rangle$, the eigenstates $\tilde \phi_i$ of $\tilde \phi(\tilde x, \tilde y)$ are linear combinations of $e^{-\langle n \rangle (1-\tilde x)/2} J_1[2 e^{-\langle n \rangle (1- \tilde x)/2}/\sqrt{p_i}]$ and $e^{-\langle n \rangle (1-\tilde x)/2} Y_1[2 e^{-\langle n \rangle (1-\tilde x)/2}/\sqrt{p_i}]$, where $p_i$ are the eigenvalues. 

\section{Single-photon subtraction}

In this Section, we present a formal derivation of Eq.\ (17) in the main text, which describes the output of a single-photon subtractor \cite{Xhoner11}.
In addition to verifying Eq.\ (17), this method allows one to treat deviations from the ideal result. 

Following Ref.\ \cite{Xhoner11}, the atoms can be in one of two collective states $|G\rangle$ and $|E\rangle$. The density matrix then evolves according to the following master equation:
\ba
\dot \rho &=& - i [\hat H_0, \rho] + \Gamma \int_0^\infty d x \Big[2 \hat \e(x) |E\rangle \langle G| \rho |G\rangle \langle E| \hat \e^\dagger(x) \nonumber \\
&& - \hat \e^\dagger(x) \hat \e(x) |G\rangle \langle G| \rho - \rho |G\rangle \langle G| \hat \e^\dagger(x) \hat \e(x)\Big].
\ea
$\hat H_0$ here describes simple propagation of light in vacuum. The photon is subtracted within a few absorption lengths $\Gamma^{-1}$ of $x = 0$, so the remainder of the medium plays no role provided $z_b > L$; hence we assumed 
$L \rightarrow \infty$. 


Here, for simplicity, we only present the derivation for two incoming photons $|2\rangle$. Generalization to an arbitrary incoming state $|\psi\rangle = \sum_n c_n |n\rangle$ is straightforward. 

Therefore, the full density matrix 
\ba
\rho = \rho_1 + |\psi_2\rangle \langle \psi_2|
\ea
 consists of the two-photon wavefunction 
\ba
|\psi_2\rangle = \frac{1}{2} \int d x d y EE(x,y) \hat \e^\dagger(x) \hat \e^\dagger(y) |0\rangle |G\rangle
\ea
and of the single-photon density matrix
\ba
\rho_1 = \int d x d y ee(x,y) \hat \e^\dagger(y) |0\rangle |E\rangle \langle E|  \langle 0| \hat \e(x).
\ea

One then finds the following equations of motion:
\ba
&& \partial_t EE(x,y) = - \partial_x EE - \partial_y EE - \Gamma  [H(x) + H(y)] EE,\nonumber \\
&& \partial_t  ee(x,y) = - \partial_x ee - \partial_y ee + 2 \Gamma \!\! \int_0^\infty \!\!\!\!\! d z EE(y,z) EE^*(x,z), \nonumber 
\ea
where $H(x)$ is the Heaviside step function. 

Starting with the boundary conditions $EE(x,y,t) = \sqrt{2} h(t-x) h(t-y)$ for $x, y \leq 0$, we solve for $EE$:
\ba
\!\!\!\!\!\!\!\!\! EE(x,y,t) = \sqrt{2} h(t-x) h(t-y) e^{-\Gamma [H(x) x + H(y) y]}.
\ea
Inserting this into the equation of motion for $ee$ and using the fact that the absorption length is much shorter than the (now uncompressed) pulse duration, we obtain
\ba
&& \partial_t ee(x,y,t) = - \partial_x ee - \partial_y ee  \nonumber \\
&&+ 2 h^2(t) h(t-x) h(t-y) [1-H(x)] [1-H(y)].
\ea
For $x, y < 0$, this can be integrated to give
\ba
ee(x,y,t) = 2 h(t-x) h(t-y) \int_{-\infty}^t h^2(t'),
\ea
so that, in the remaining three quadrants of the $xy$ plane,
\ba
\!\!\!\!\!\!\!\!\!\!\!\! ee(x,y,t) = 2 h(t-x) h(t-y) \int_{-\infty}^{t-\textrm{max}(x,y)} d t' h^2(t'), 
\ea
which is a special case of Eq.\ (17) in the main text.


%

\end{document}